\documentclass[12pt,eqsecnum]{article}
\usepackage[dvips]{graphicx}
\usepackage{amssymb}
\usepackage{amsmath}
\usepackage{epstopdf}
\makeatletter

\@addtoreset{equation}{section}
\makeatletter

\newcommand{\ma}[1]{\mbox{$\mathcal{#1}$}}

\newcommand{\D}{{\rm d}}

\newcommand{\dalm}{\kern1pt\vbox{\hrule height 0.9pt\hbox{\vrule width
0.9pt\hskip 2.5pt\vbox{\vskip 5.5pt}\hskip 3pt\vrule width 0.3pt}\hrule height
0.3pt}\kern1pt}

\def\b2hat{ {\hat b}_2 }

\newcommand{\Ri}{{\cal R}}
\def\none{\nonumber\\}

\def\be {\begin{equation}}
\def\ee  {\end{equation}}
\def\bea {\begin{eqnarray}}
\def\eea {\end{eqnarray}}

\begin{document}

\begin{titlepage}
\vfill
\begin{flushright}
\today
\end{flushright}

\vfill
\begin{center}
\baselineskip=16pt
{\Large\bf 
Geometrodynamics of spherically symmetric\\
Lovelock gravity\\
}
\vskip 0.5cm
{\large {\sl }}
\vskip 10.mm
{\bf Gabor Kunstatter${}^{a}$, Tim Taves${}^{a}$, and Hideki Maeda${}^{b}$} \\

\vskip 1cm
{
	${}^a$ Department of Physics, University of Winnipeg and Winnipeg Institute for Theoretical Physics, Winnipeg, Manitoba,Canada R3B 2E9\\
	${}^b$ Centro de Estudios Cient\'{\i}ficos (CECs), Casilla 1469, Valdivia, Chile. \\
	\texttt{g.kunstatter@uwinnipeg.ca, timtaves@gmail.com, hideki@cecs.cl}

     }
\vspace{6pt}
\today
\end{center}
\vskip 0.2in
\par
\begin{center}
{\bf Abstract}
 \end{center}
\begin{quote}
We derive the Hamiltonian for spherically symmetric Lovelock gravity using the geometrodynamics approach pioneered by Kucha\v{r}~\cite{kuchar94} in the context of four-dimensional general relativity. When written in terms of the areal radius, the generalized Misner-Sharp mass and their conjugate momenta, the generic Lovelock action and Hamiltonian take on precisely the same simple forms as in general relativity. This result supports the interpretation of Lovelock gravity as the natural higher-dimensional extension of general relativity. It also provides an important first step towards the study of the quantum mechanics, Hamiltonian thermodynamics and formation of generic Lovelock black holes. 
  \vfill
\vskip 2.mm
\end{quote}
\end{titlepage}





\newpage

\section{Introduction}

The elegance of its underlying geometrical structure makes Einstein's general theory of relativity both beautiful and compelling. More importantly, observational consequences of general relativity have been verified over a large range of macroscopic scales, from terrestial to cosmological. The inevitability of singularities in the context of cosmology and black holes makes it clear that  modifications to the Einstein-Hilbert action are inevitable at very short distance scales. Quantum theory suggests the need to add higher-curvature terms whereas string/M-theory~\cite{string} requires the existence of extra spatial dimensions. The most natural such generalization of the Einstein-Hilbert action in $n$ dimensions { that is quasi-linear second-order} is provided by so-called Lovelock gravity~\cite{Lovelock}, which consists of a sum of $[n/2]$ dimensionally extended Euler densities that are non-trivial only in dimensions greater than four. ($[n/2]$ refers to the largest integer less than or equal to $n/2$.) 

The fact that Lovelock gravity is second-order implies that it is ghost-free when linearized around a flat background\cite{ghost free}. Moreover, string theoretic arguments~\cite{string lovelock} suggest that { quadratic} Lovelock gravity appears in the low-energy limit for strings propagating in curved spacetime. 
Since Lovelock gravity is a natural extension of general relativity in arbitrary dimensions, it could potentially provide insights into quantum gravity or even a hint concerning the origin of the four-dimensionality of our universe.
Lovelock gravity has therefore received considerable attention recently leading to interesting developments at the classical level. {(See~\cite{lovelockreview} for review.)} There are, however, few results on its quantum aspects. The purpose of this letter is to pave the way towards a systematic study of the quantization of Lovelock gravity.

At first glance, the action for Lovelock gravity appears quite complicated~\cite{Lovelock}:
\bea
I&=&\frac{1}{2\kappa_n^2}\int d^nx\sqrt{-g}\sum^{[n/2]}_{p=0}\alpha_{(p)} {\cal L}_{(p)},
\label{eq:lovelock action}\\
{\cal L}_{(p)}&:=& \frac{p!}{2^p}\delta^{\mu_1...\mu_p\nu_1..\nu_p}_{\rho_1...\rho_p\sigma_1..\sigma_p}
    R_{\mu_1\nu_1}{}^{\rho_1\sigma_1}...R_{\mu_p\nu_p}{}^{\rho_p\sigma_p},
 \label{eq:lovelock lagrangian}
\eea
where
$
\delta^{\mu_1...\mu_p}_{\rho_1...\rho_p}
:=
\delta^{\mu_1}_{[p_1}...\delta^{\mu_p}_{p_p]}\,
$.
Each $\alpha_{(p)}$ is a coupling constant of dimension (length) ${}^{2(p-1)}$. { $\alpha_{(0)}$ is proportional to the cosmological constant} and the $p=1$ term corresponds to { general relativity}.
As stated above, the specific combination of curvature terms in each of the ${\cal L}_{(p)}$ guarantees that the equations for Lovelock gravity are second order in the metric components and ghost-free\cite{ghost free}. Equally important is that the theory obeys a generalized Birkhoff theorem~\cite{gen birkhoff,HM11}: for a given set of coupling constants which do not admit degenerate vacua, all $C^2$-class spherically symmetric vacuum solutions except for the Nariai-type solution admit a hypersurface orthogonal Killing vector and can be parametrized by a single constant of motion. 
The spherically symmetric vacuum solutions have been fairly well studied~\cite{lovelockreview}, but little work has been done on the quantum mechanics perhaps because of the perceived complexity of the equations. 

The Hamiltonian analysis  for full Lovelock gravity, first considered by Teitelboim and Zanelli~\cite{TZ}, is qualitatively different from that of general relativity due to the higher-order nature of the theory~\cite{TZ,df2011}.
For the case of spherical symmetry, the geometrodynamics of five-dimensional Einstein-Gauss-Bonnet (i.e. $p=2$ Lovelock gravity) was worked out by Louko {\it et al}~\cite{JL97}, while the Hamiltonian analysis of higher-dimensional Gauss-Bonnet coupled to matter was recently done in~\cite{TLKM}.

In the following, we will show that the geometrodynamic formulation of Kucha\v{r}~\cite{kuchar94} yields a Hamiltonian for spherically symmetric $p$-th order Lovelock gravity that is just as simple and elegant as that of Einstein's original theory. This result suggests that Lovelock gravity is in a deep sense a natural higher-dimensional extension of general relativity. It also provides a first important step towards a more complete understanding of the quantum properties of Lovelock black holes.


\section{Geometrodynamics of Lovelock Black Holes}
We start with a general diffeomorphism invariant theory of gravity that obeys the Birkhoff theorem {in  spherically symmetric spacetimes.
Since an $n(\ge 4)$-dimensional spherically symmetric spacetime $({\ma M}^n, g_{\mu \nu })$ is a warped product of an 
$(n-2)$-dimensional unit maximally symmetric space $(S^{n-2}, \gamma _{ab})$ with positive sectional curvature~\footnote{The discussion can easily be generalized to included zero or negative sectional curvature\cite{KMT12}.} and a two-dimensional orbit spacetime $(M^2, g_{AB})$, the line element may be given by
\begin{align}
g_{\mu \nu }\D x^\mu \D x^\nu =g_{AB}({\bar y})\D {\bar y}^A\D {\bar y}^B +R^2({\bar y}) d\Omega^2_{n-2},
\label{eq:ansatz}
\end{align} 
where $d\Omega^2_{n-2}$ is the line element on $(S^{n-2}, \gamma _{ab})$, and the areal radius $R$ is a scalar on $(M^2, g_{AB})$. 
In a spherically symmetric spacetime, the system may be described by the effective two-dimensional action; 
\bea
I_{(2)}=\int d{\bar y}^0L[g_{AB},R]=\int d{\bar y}^0 \int d{\bar y}^1{\cal L}[g_{AB},R],\label{2-action}
\eea
where ${\bar y}^0$ is a timelike coordinate on $(M^2, g_{AB})$.
Here the Lagrangian $L$ and the Lagrangian density ${\cal L}$ are functionals of the metric functions and their derivatives, which are determined up to a total derivative.
The main purpose of geometrodynamics is to find canonical variables (that are functionals of the metric functions) to provide a tractable form and transparent physical meaning for ${\cal L}$.

If $(D R)^2:=g^{AB}(D_AR)(D_BR) \ne 0$, where $D_A$ is the covariant derivative on $(M^2, g_{AB})$, the metric may be written in terms of the areal coordinates as 
\be
ds^2 = -F(R,T)e^{2\sigma(R,T)}dT^2 + F^{-1}(R,T) dR^2 + R^2 d\Omega^2_{n-2}.
\label{eq:schwarz metric}
\ee
The original Birkhoff theorem in general relativity asserts that the spherically symmetric vacuum solution is uniquely given by the Schwarzschild-Tangherlini solution, which is given in the coordinates (\ref{eq:schwarz metric}) by $F(R,T)=F(R;M)$ and $\sigma(R,T)=0$, where $M$ has the interpretation as the mass and $T$ is the so-called Schwarzschild time.
In the presence of a cosmological constant, the Nariai solution is also possible where $(D R)^2=0$ is satisfied.
In Lovelock gravity, the most relevant part of the statement of the Birkhoff theorem is that the $C^2$ vacuum solution with $(D R)^2 \ne 0$ is the Schwarzschild-Tangherlini-type solution if the theory does not admit degenerate vacua~\cite{gen birkhoff,HM11}.
(The theorem in the quadratic theory was proven in~\cite{cd2002,HM08}.)
{ We assume that the metric is $C^2$, keeping in mind that there are counterexamples in the literature\cite{Garraffo:2007fi}. }

In the fully dynamical setting, we can without loss of generality consider precisely the same form of metric but with $M=M(R,T)$ and $\sigma=\sigma(R,T)$ both as functions of $R$ and $T$. In that case $M(R,T)$ has the geometrical/physical interpretation as the generalized Misner-Sharp mass function. 
Because of $(D R)^2=F(R;M(R,T))$, the mass function can in principle be expressed as a function of $(D R)^2$, as long as $(DR)^2 \ne 0$, the theory which does not admit degenerate maximally symmetric vacua and one knows the form of the static spherically symmetric solution, i.e. $F(R,M)$. 
In Lovelock gravity, the mass function is given by
\bea
M&:=& \frac{(n-2)A_{n-2}}{2\kappa_n^2}\sum_{p=0}^{[n/2]} \tilde{\alpha}_{(p)}R^{n-1-2p}\left(1-(D R)^2\right)^p\none
&=& \frac{(n-2)A_{n-2}}{2\kappa_n^2}\sum_{p=0}^{[n/2]} \tilde{\alpha}_{(p)}R^{n-1-2p}\left(1-F(R,M)\right)^p,
\label{eq:M EL}
\eea
where $A_{n-2}$ is the area of $(S^{n-2}, \gamma _{ab})$ and ${\tilde \alpha}_{(p)}:=(n-3)!\alpha_{(p)}/(n-1-2p)!$~\cite{misner_Sharp,maeda2006,HM08,HM11}.
In $n$-dimensional general relativity without cosmological constant, it reduces to 
\be
M= \frac{(n-2)A_{n-2}}{2\kappa_n^2}\alpha_{(1)}R^{n-3}\left(1-(D R)^2\right),
\label{eq:M GR}
\ee
and hence
\be
F(R,M)= 1-\frac{2\kappa_n^2 M}{(n-2)A_{n-2}\alpha_{(1)}R^{n-3}}.
\label{eq:F GR}
\ee
{ In general, the expression for $F(R,M)$ is more complicated. In order for the theory to provide a unique solution for each $M$, the relationship between $M$ and $F$ must be monotonic, which provides a condition on the Lovelock coupling constants $\alpha_{(p)}$ in (\ref{eq:M EL}). We will for what follows assume that this condition is satisfied, but will return to the more general case in \cite{KMT12}.}

The essence of Kucha\v{r}' geometrodynamics performed in vacuum four-dimensional general relativity is to write the Hamiltonian equations describing the dynamics of a general spherically symmetric geometry in terms of the areal radius, mass function and their conjugate momenta. We now present the geometrodynamics of spherically symmetric Lovelock gravity.

For spherically symmetric spacetimes, the action reduces to
\bea
I=\frac{A_{n-2}}{2\kappa_n^2}\int d^2{\bar y}\sqrt{-g_{(2)}}R^{n-2}\sum^{[n/2]}_{p=0}\alpha_{(p)} {\cal L}_{(p)},
\eea
where $g_{(2)}:=\det(g_{AB})$ and the dimensionally reduced $p$th order Lovelock term is given from expressions (2.19) and (2.20) of~\cite{HM11} as 
\begin{align}
{\cal L}_{(p)} =& \frac{(n-2)!}{(n-2p)!} \Biggl[(n-2p)(n-2p-1)\left(\frac{1-(DR)^2}{R^2}\right)^{p} - 2p(n-2p)\frac{D^2R}{R}\left(\frac{1-(DR)^2}{R^2}\right)^{p-1} \none
& + 2p(p-1) \frac{(D^2R)^2 - (D^AD_BR)(D^BD_AR)}{R^2} \left(\frac{1-(DR)^2}{R^2}\right)^{p-2} + p\overset{(2)}{\Ri} \left(\frac{1-(DR)^2}{R^2}\right)^{p-1} \Biggr],  \label{L_p}
\end{align}
where ${}^{(2)}\Ri$ is the Ricci scalar on $(M^2, g_{AB})$ and $D^2R:=D^AD_AR$.
The contraction was taken over the two-dimensional orbit space.
Using the binomial expansion and collecting terms with the same number of derivatives yields
\begin{align}
\label{eq:lp1}
{\cal L}_{(p)} =& \frac{(n-2)!}{(n-2p)!} \Biggl[(n-2p)(n-2p-1)\left(\frac{1-(DR)^2}{R^2}\right)^{p} -2p(n-2p)\frac{D^2R}{R}\left(\frac{1-(DR)^2}{R^2}\right)^{p-1} \none 
& + pR^{2-2p}\overset{(2)}{\Ri}  + \displaystyle\sum\limits_{i=0}^{p-2} \frac{2(-1)^ip! (DR)^{2i}}{(i+1)!(p-2-i)!} \biggl\{(i+1)\frac{(D^2R)^2 - (D^AD_BR)(D^BD_AR)}{R^{2p-2}} \none
&- \frac{D^AR(D^2D_AR - D_AD^2R) (DR)^2}{R^{2p-2}} \biggl\}\Biggr],  
\end{align}
where we have removed some terms containing  ${}^{(2)}\Ri$ by using the two-dimensional identity:
\be
(D R)^2 \overset{(2)}{\Ri}\equiv  2(D^A R)\left(D^2D_AR - D_AD^2R\right). 
\ee
This identity can be derived from Eq.~(2.10) of \cite{TLKM}.

Using integration by parts, the term inside curly brackets in (\ref{eq:lp1}) can be replaced by the following remarkably simple term:
\be
\sum\limits_{i=0}^{p-2} \frac{2(-1)^ip!}{(i+1)!(p-2-i)!}  D_A(R^{n-2p})((DR)^2)^i\biggl\{\frac12 D^A((DR)^2) - (D^AR)(D^2R) \biggl\}\, .  
\ee
Finally, the $p$th order Lovelock Lagrangian is, up to total divergences: 
\begin{align}
\label{eq:lovelock simplified}
 {\cal L}_{(p)} =& \frac{(n-2)!}{(n-2p)!} \Biggl[p\overset{(2)}{\Ri} R^{2-2p} + pR^{2-n}\frac{D^A(R^{n-2p})D_A((DR)^2)}{(DR)^2}\biggl\{1-(1-(DR)^2)^{p-1}\biggl\}  \none
& + (n-2p)(n-2p-1)\biggl\{\left(1-(DR)^2\right)^{p} +2p(DR)^2\biggl\}R^{-2p} \Biggr]. 
\end{align}
This is the first key result of the present paper. 
{ Its value becomes evident when considering the following Arnowitt-Deser-Misner (ADM) decomposition on $(M^2, g_{AB})$:}
\be
ds^2= -N(t,x)^2dt^2+\Lambda(t,x)^2(dx+N_r(t,x)dt)^2+R^2(x,t)d\Omega_{n-2}^2.
\label{eq:kuchar adm}
\ee
Specifically, Eq.~(\ref{eq:lovelock simplified}) enables us to find a canonical transformation from ADM phase space variables $(\Lambda,P_\Lambda;R,P_R)$  to $(M,P_M;S,P_S)$.
(Our $x$, $S$, and $P_S$ are equivalent to Kucha\v{r}'s $r$, ${\rm R}$ and $P_{\rm R}$, respectively.) The new canonical variable $S$ is then chosen as $S=R$ and hence also represents the areal radius.

$(DR)^2$ contains no derivatives of $\Lambda$, while the Ricci scalar term in the action takes the form:
 \bea
 \sqrt{-g_{(2)}}R^{n-2p}\overset{(2)}{\Ri} &=&-2N^{-1}\biggl\{(R^{n-2p})\dot{} - N_r (R^{n-2p})'\biggl\}(\dot{\Lambda} -N'_r\Lambda-N_r\Lambda')\none
 & & -2N\biggl\{(R^{n-2p})''\Lambda^{-1}+(R^{n-2p})'(\Lambda^{-1})'\biggl\} + \partial_t(...) + \partial_x(...), 
 \label{eq:Ricci}
 \eea
{where a dot and a prime denote the partial derivative with respect to $t$ and $x$, respectively.}
Thus it is evident that the generic Lovelock term is linear in $\dot{\Lambda}$, a fact that plays a crucial role in the following analysis. We therefore write the effective two-dimensional Lagrangian density ${\cal L}$ in (\ref{2-action}) as
\be
{\cal L}= B_0(R,y,\Lambda)\dot{\Lambda}+ B_1(R,y,\Lambda),
\label{eq:LG}
\ee
where we have replaced $\dot{R}$ in the above by
\be
y:= \frac{1}{N}\left(\dot{R}- N_r R'\right) 
\label{eq:y}
\ee
because general covariance of the action guarantees that $\dot{R}$ always appears in this combination~\cite{kuchar94}.
Equation~(\ref{eq:LG}) implies that the conjugate to $\Lambda$ is
\be
P_\Lambda=B_0(R,y,\Lambda),
\label{eq:PLambda}
\ee
which implicitly { determines $y=y(\Lambda,P_\Lambda,R)$ as a function of $(\Lambda, P_\Lambda, R$)}. It is important for the following that $y$ is independent of $P_R$, but the explicit form of $B_0(R,y,\Lambda)$ is not crucial.

Following~\cite{kuchar94}, we define a new set of variables $(M,P_M,S,P_S)$ from the ADM phase space variables $(\Lambda,P_\Lambda,R,P_R)$ by
\bea
P_M&:=&-e^{\sigma} T'=-\frac{y\Lambda}{F},  
\label{eq:PM}\\
S&:=&R,
\label{eq:R}\\
P_S &:=& P_S(\Lambda,P_\Lambda,R) = \frac{1}{R'} ( P_R R'-\Lambda P_\Lambda' - P_M M'), 
\label{eq:Pr1}
\eea 
while $M$ is given as a function of $(\Lambda,P_\Lambda,R)$ by solving Eq.~(\ref{eq:M EL}) and 
\bea
F(M,R) = \frac{R'^2}{\Lambda^2} - y^2.  
\label{eq:F}
\eea 
We recall $y=y(\Lambda,P_\Lambda,R)$ as implied by (\ref{eq:PLambda}).

An explicit calculation reveals that this transformation preserves the Liouville form up to a total variation and is therefore canonical:
\be
P_\Lambda\delta{\Lambda}+P_R\delta{R} =P_M\delta{M}+P_S\delta{S} + \delta(...).  
\ee 
As argued by Kucha\v{r}~\cite{kuchar94} the specific form of $P_S$ is determined by the transformation properties of the variables under spatial diffeomorphisms. Equations~(\ref{eq:PM})--(\ref{eq:F}) are completely generic in form. The theory specific information is contained in the functions $F(M,R)$ and $y(\Lambda,P_\Lambda,R)$.

To proceed, we note the key identity:
\bea
& &\frac{D^A(R^{n-2p})D_A((DR)^2)}{(DR)^2}\{1-(1-(DR)^2)^{p-1}\} \none 
 &=&(n-2p)\frac{R^{n-2p-1}}{FN\Lambda}\{1-(1-F)^{p-1}\}\biggl\{-\Lambda y{\dot F}+\left(\frac{NR'}{\Lambda} + N_r\Lambda y\right)F'\biggl\}.
\label{eq:observation1}
\eea
By substituting (\ref{eq:observation1}) and (\ref{eq:Ricci}) into (\ref{eq:lovelock simplified}) one obtains:
\begin{align}
{\cal L}_{(p)}=& \frac{(n-2)!}{(n-2p-1)!}pN^{-1}\Lambda^{-1} R^{1-2p}\biggl[2y(N_r'\Lambda +N_r\Lambda')  \nonumber \\
&-\frac{2N}{(n-2p)R^{n-2p-1}}\biggl\{(R^{n-2p})''\Lambda^{-1}+(R^{n-2p})'(\Lambda^{-1})'\biggl\}\nonumber \\
&+ {F}^{-1}\biggl\{1-(1-{F})^{p-1}\biggl\}(\Lambda N_r y+\Lambda^{-1}NR')F' -\biggl(2y{\dot \Lambda} + \Lambda y\frac{\dot F}{F}\biggl) +(1-{F})^{p-1}\Lambda y\frac{\dot F}{F} \biggl]. \label{eq:lp2}
\end{align}
Eqs.~(\ref{eq:M EL}), (\ref{eq:PM})  and (\ref{eq:F}) imply that:
\bea
P_M\dot{M} &=&\frac{(n-2)A_{n-2}}{2\kappa_n^2}\sum_{p=0}^{[n/2]}\tilde{\alpha}_{(p)}R^{n-1-2p}\frac{y\Lambda}{F}\left\{p(1-F)^{p-1}\dot F-\frac{n-1-2p}{R}(1-F)^p\dot R\right\}.\none
\label{eq:pm1}
\eea
The result for $P_M M'$ is analoguous with dots replaced by primes.

Using the above expressions for $P_M\dot{M}$ and $P_M M'$ as well as (\ref{eq:lp2}) it is straightforward, but algebraically lengthy, to confirm the following via direct substition:
\bea
{\cal L}-P_M\dot{M} - \frac{N\Lambda}{R'} M' &=&\frac{y\Lambda }{F}\left(N_r + N \frac{y}{R'}\right)M'\none
 &=&-\frac{{\dot R}}{R'}P_MM'. 
\label{eq:Lfinal}
\eea

From Eq.~(\ref{eq:Lfinal}) it follows immediately that the Hamiltonian in the equivalent two-dimensional theory takes the form:
\bea
{\cal H}_G&:=& P_M\dot{M} + P_S\dot{S}-{\cal L} \none
  &=& - \frac{\Lambda}{R'}\biggl(N+\frac{y{\dot R}}{F}\biggl)M' + P_S\dot{S}  \none
   &=& N^M M'+N^S P_S,
   \label{eq:geometrodynamic hamiltonian}
   \eea
where { we have used Eq.~(\ref{eq:Lfinal}) to get the second line} and  defined new Lagrange multipliers $N^M$ and $N^S$. 
{ The Lagrangian density for the canonical coordinates $(M,S,N^M,N^S)$ can now be written as}
\bea
{\cal L}= P_M\dot{M} + P_S\dot{S}- N^M M'-N^S P_S. \label{eq:geometrodynamic L}
   \eea
Eq.~(\ref{eq:geometrodynamic L}) is precisely the same form as Eq.~(122) of \cite{kuchar94}. { In comparison to the rather complicated starting point in Eq.~(\ref{L_p}), this equivalent Lagrangian density is extremely simple and the physical meaning of the canonical variables are very clear.} This is our main result. Remarkably, the coupling constants $\alpha_{(p)}$ do not appear explicitly in any of the equations after (\ref{eq:PM}).
They are in fact hidden in the definition of the mass function. This makes it possible to treat any class of Lovelock gravity in exactly the same way.
 
The constraints, $M'=0$ and $P_S=0$, { are obtained by varying the Lagrange multipliers $N^M$ and $N^S$, respectively}. On the constraint surface $M=m(t)$, as expected, and $P_S=0$.  The reduced phase space is therefore two-dimensional consisting of $p_m:=\int^\infty_{-\infty} dx P_M(x,t)$ and $m$.
With suitable boundary conditions~\cite{KMT12}, one can repeat the analysis of~\cite{kuchar94} for spacelike slicings that intersect both left and right branches of the outer horizons of eternal black holes to obtain the reduced action:
\be
I_{(2)} = \int dt \biggl[p_m\dot{m} - ({N}_+-{N}_-)m\biggl],
\label{eq:reduced action}
\ee
{ where $N_{\pm}:=\mp \lim_{x\to \pm\infty}N^M$. The reduced equations of motion in vacuum then imply that $m=m_0=constant$, and $\dot{p}_m = -(N_+-N_-)$. }

\section{Conclusions}

{
In summary, we have presented the results of a geometrodynamics/Hamiltonian analysis of spherically symmetric Lovelock gravity. 
The equivalent two-dimensional Lagrangian density (\ref{eq:geometrodynamic L}) and Hamiltonian (\ref{eq:geometrodynamic hamiltonian}) written in terms of the areal radius, the generalized Misner-Sharp mass, and their conjugate momenta have exactly the same simple forms as in general relativity.
Our analysis paves the way for further understanding of generic Lovelock black holes. Details of the proof will be presented in the subsequent paper~\cite{KMT12}. 
}


Given (\ref{eq:reduced action}), the reduced quantization of Lovelock black holes can be performed exactly as in \cite{kuchar94}. Alternatively one can take the approach of Louko and M\"akel\"a~\cite{LM96} to derive the quantization of the throat of the Lovelock black hole to obtain the mass spectrum as a function of { the number of spacetime dimensions} and Lovelock couplings. This is currently in progress.
As well, the geometrodynamics described above is the starting point for the Hamiltonian thermodynamics of Lovelock black holes, following \cite{JL97}.  Finally, we remark that, as we will show in \cite{KMT12}, the Hamiltonian (\ref{eq:geometrodynamic hamiltonian}) allows one to write down the reduced  Hamiltonian equations for black-hole formation via scalar-field collapse, either in generalized Painlev\'{e}-Gullstrand coordinates, as done for Einstein-Gauss-Bonnet gravity in~\cite{TLKM} or using the local Hamiltonian of~\cite{pullin11,GK11}. This in turn provides the starting point for a detailed numerical study of black-hole formation in Lovelock gravity. \\[20pt]

{\bf Acknowledgments}
This research was supported in part by the Natural Sciences and Engineering Research Council of Canada and by the JSPS Grant-in-Aid for Scientific Research (A) (22244030). This work has been partially funded by the Fondecyt grants 1100328, 1100755 and by the Conicyt grant "Southern Theoretical Physics Laboratory" ACT-91. 
The Centro de Estudios Cient\'{\i}ficos (CECs) is funded by the Chilean Government through the Centers of Excellence Base Financing Program of Conicyt. HM thanks Jorge Zanelli for valuable comments. GK and TT are very greatful to CECs, Chile for its kind hospitality during the initial stages of this work. GK also thanks Jorma Louko for introducing him to the beauty of Kucha\v{r}' geometrodynamics. TT and GK also thank Danielle Leonard and Robert Mann for the collaboration in \cite{TLKM} which was instrumental in leading to the present work.

\end{document}